\documentstyle[twocolumn]{article}

\newlength{\spacing}
\newcommand{\doublespace}{\setlength{\baselineskip}{1.5\spacing}}
\newtheorem{thm}{Theorem}[section]

\newtheorem{lem}[thm]{Lemma}

\def\rar{\to}

\def\today{\ifcase\month\or
  January\or February\or March\or April\or May\or June\or
  July\or August\or September\or October\or November\or December\fi
  \space\number\day, \number\year}

\begin{document}
\begin{center}
%
%
{\bf Comment to ``Coverage by Randomly Deployed Wireless Sensor Networks".} \\
\vspace{0.20in} by \\
\vspace{0.1in} {Bhupendra Gupta \footnote{Corresponding Author, email: bhupen@iiitdm.in, gupta.bhupendra@gmail.com}}\\
Faculty of Engineering and Sciences,\\
Indian Institute of Information Technology (DM)-Jabalpur-482011, India.\\
\vspace{0.1in}
\end{center}
\doublespace
\section{Introduction.}
In the above paper, Lemma (4), on page $2659$ play the key role for
deriving the main results in the paper. The statement as well as the
proof of Lemma (4), page $2659,$ \cite{wan} is not correct. Here
mainly two serious errors in this Lemma, one is in the statement of
lemma and the other one in the proof of Lemma (4), on page 2666,
\cite{wan}.
In the Lemma (4), page 2659, \cite{wan}, Wan and Yi, states the
lemma as follows,
\begin{lem}
Assume $\Omega$ is the disk region and let $r_n$ be such that $n\pi
r_n^2 \rar \infty$ and $n \pi r_n^3 \rar 0.$ Then for any point $z
\in \delta \Omega$
\begin{equation}
\phi_{n,r_n}(z) \sim \frac{\left(\frac{n\pi
r_n^2}{2}\right)^k}{k!}e^{-\frac{n\pi r_n^2}{2}}.\label{a}
\end{equation}
\label{lemma}
\end{lem}
Firstly the statement of Lemma (4), page 2659, \cite{wan}, is wrong.
Since authors has taken $n\pi r_n^2 \rar \infty.$ This implies
\[\phi_{n,r_n}(z) \rar 0,\]
because in the right hand side of (\ref{a}), $n\pi r_n^2$ is in the
negative power of exponential.\\
Hence authors should remove the condition $n\pi r_n^2 \rar \infty,$
from the statement of Lemma (\ref{lemma}) (Lemma (4), page 2659,
\cite{wan}).\\

Now the second and more serious error is in the proof of Lemma (4),
on page 2666, \cite{wan}. Here authors has derive the lower bound of
$\phi(z)$ as follows
\begin{equation}
\phi(z) \geq \sum_{i=0}^{k}\frac{\left(\frac{n\pi
r_n^2}{2}\right)^i}{i!}e^{-\frac{n\pi r_n^2}{2}} >
\frac{\left(\frac{n\pi r_n^2}{2}\right)^k}{k!}e^{-\frac{n\pi
r_n^2}{2}}.\label{b}
\end{equation}
By using the second inequality, authors were getting lose lower
bound, while the objective of the lemma is to get the exact
expression for $\phi(z).$\\

Also, authors has given the proof for the upper bound of $\phi(z)$
as follows
\begin{eqnarray*}
\phi(z) & \leq & \sum_{i=0}^{k} \frac{\left(\frac{n\pi
r_n^2}{2}\right)^i}{i!}e^{-\frac{n\pi r_n^2}{2}+ nr^2 arcsin
\frac{\sqrt{\pi}r}{2}}\\
& = & \left(\sum_{i=0}^{k}\frac{\left(\frac{n\pi
r_n^2}{2}\right)^i}{i!}e^{-\frac{n\pi r_n^2}{2}}\right)e^{nr^2
arcsin \frac{\sqrt{\pi}r}{2}}\\
& \sim & \sum_{i=0}^{k}\frac{\left(\frac{n\pi
r_n^2}{2}\right)^i}{i!}e^{-\frac{n\pi r_n^2}{2}} \sim
\frac{\left(\frac{n\pi r_n^2}{2}\right)^k}{k!}e^{-\frac{n\pi
r_n^2}{2}}
\end{eqnarray*}
The problem is with the last `$\sim$' in the above expression. From
(\ref{b}), we have
\[\sum_{i=0}^{k}\frac{\left(\frac{n\pi
r_n^2}{2}\right)^i}{i!}e^{-\frac{n\pi r_n^2}{2}} >
\frac{\left(\frac{n\pi r_n^2}{2}\right)^k}{k!}e^{-\frac{n\pi
r_n^2}{2}}.\]
It is clear that one can not replace `$a > b$' by `$a \sim b$.' Even
in case of clear convergence i.e., $a \rar b,$ here it is not
allowed, because the authors were giving the derivation of upper
bound for $\phi(z)$ and $\frac{\left(\frac{n\pi
r_n^2}{2}\right)^k}{k!}e^{-\frac{n\pi r_n^2}{2}}$ is the lower bound
of $\sum_{i=0}^{k}\frac{\left(\frac{n\pi
r_n^2}{2}\right)^i}{i!}e^{-\frac{n\pi r_n^2}{2}}.$\\

Following is the corrected version of Lemma (4).
\begin{lem}
Assume $\Omega$ is the disk region and let $r_n$ be such that $n \pi
r_n^3 \rar 0.$ Then for any point $z \in \delta \Omega$
\begin{equation}
\sum_{i=1}^{k}\frac{\left(\frac{n\pi
r_n^2}{2}\right)^i}{i!}e^{-\frac{n\pi r_n^2}{2}} \leq
\phi_{n,r_n}(z) \leq (1+C) \sum_{i=1}^{k}\frac{\left(\frac{n\pi
r_n^2}{2}\right)^i}{i!}e^{-\frac{n\pi r_n^2}{2}},\label{a1}
\end{equation}
where $C>0$ is an arbitrary constant.
\label{lemma}
\end{lem}
\textbf{Proof.} The idea of the proof is same, except the
mathematics for getting the bounds of $\phi(z),$ which leads to
totally different bounds. Here we are giving only those
steps which are different from the original proof.\\
For the lower bound
\[\phi(z) \geq \sum_{i=1}^{k}\frac{\left(\frac{n\pi
r_n^2}{2}\right)^i}{i!}e^{-\frac{n\pi r_n^2}{2}}.\]
Now for the upper bound
\begin{eqnarray*}
\phi(z) & \leq & \sum_{i=0}^{k} \frac{\left(\frac{n\pi
r_n^2}{2}\right)^i}{i!}e^{-\frac{n\pi r_n^2}{2}+ nr^2 arcsin
\frac{\sqrt{\pi}r}{2}}\\
& = & \left(\sum_{i=0}^{k}\frac{\left(\frac{n\pi
r_n^2}{2}\right)^i}{i!}e^{-\frac{n\pi r_n^2}{2}}\right)e^{nr^2
arcsin \frac{\sqrt{\pi}r}{2}}\\
& \leq & (1+C)\sum_{i=0}^{k}\frac{\left(\frac{n\pi
r_n^2}{2}\right)^i}{i!}e^{-\frac{n\pi r_n^2}{2}},
\end{eqnarray*}
since $nr^2 acrsin \frac{\sqrt{\pi}r}{2} = \frac{1}{2}n\pi^{1/2}r^3
+ \frac{1}{2^3 6}n\pi^{3/2}r^5+\ldots,$ and $n \pi r_n^3 \rar 0.$\\

This change in Lemma leads a drastic change in all the result
derived in the article \cite{wan}.

\end{document}